\documentclass[twocolumn,eqsecnum,showpacs,preprintnumbers,amssymb,aps]{revtex4}
\usepackage{graphicx}
\usepackage{dcolumn}
\usepackage{bm}

\begin{document}

\title{Spectacular Role of Electron Correlation in the Hyperfine Interactions in $^2D_{5/2}$ States in Alkaline Earth Ions} 
\vspace{0.5cm}

\author{$^{1,2}$B. K. Sahoo, $^{2}$C. Sur, $^{1}$T. Beier, $^{2}$B. P. Das, $^{2}$R. K. Chaudhuri and $^{3}$D. Mukherjee \\
\vspace{0.3cm}
$^{1}${\it Atomphysik, GSI, Planckstra$\beta$e 1, 64291 Darmstadt, Germany}\\
$^{2}${\it NAPP Group, Indian Institute of
Astrophysics, Bangalore-34, India}\\
$^{3}${\it Department of Physical Chemistry, IACS, Kolkata-700 032, India}}
\date{\today}
\vskip1.3cm
\begin{abstract}
\noindent
The low-lying n(=3,4,5)d $^2D_{5/2}$ states alkaline earth ions are of vital importance in a number of different physical applications. The hyperfine structure constants of these states are characterized by unusually strong electron correlation effects. Relativistic coupled-cluster theory has been employed to carry out {\it ab initio} calculations of these constants. The role of the all order core-polarization effects was found to be decisive in obtaining good agreement of the results of our calculations with accurate measurements. The present work is an apt demonstration of the power of the coupled-cluster method to cope with strongly interacting configurations.
\end{abstract} 
\maketitle

The $^2S_{1/2} \rightarrow {^2D}_{5/2}$ transition frequencies in singly ionized alkaline earth atoms are among the leading candidates in the search for an optical frequency standard \cite{champenois,barwood,sherman}. They are also of interest from the point of view of quantum information processing \cite{roos}, parity non-conservation arising from the nuclear anapole moment \cite{geetha}, and astrophysics \cite{beiersdorfer}. A knowledge of the magnetic dipole hyperfine structure constant ($A$) for the $^2D_{5/2}$ states of these systems is essential for determining frequency standards \cite{itanof}. Hyperfine interactions in these states are strongly influenced by electron correlation and it is therefore of crucial importance to use a method that can treat accurately the strongly interacting configurations. The primary objective of this paper is to study the peculiar trends in the electron correlation in the hyperfine interaction constants for the $^2D_{5/2}$ states of the singly ionized alkaline earth atoms using the relativistic coupled-cluster (RCC) theory. This theory is equivalent to all-order many-body perturbation theory \cite{lindgren,bijaya3} and it has the potential to take into account the interplay of relativistic and correlation effects. It has been successfully applied to understand the role of electron correlations in $s$ and $p$ valence systems \cite{bijayaba,bijaya2,bijayapb}. However, the excited $^2D_{5/2}$ states of the alkaline earth ions are strongly affected by the core electrons. Systematic investigations of the core-polarization (CP) effects in these systems can be carried out to obtain important insights into these phenomena.

The $A$ constant is given by \cite{cheng}
\begin{eqnarray}
A = \mu_N \ g_I \ \frac {\langle J || \textbf{T}^{(1)}||J\rangle}{\sqrt{J(J+1)(2J+1)}}, \label{eqn1}
\end{eqnarray}
where $g_I= [\frac {\mu_I}{I}]$ is the nuclear Land\a'e g-factor with $\mu_I$ and $\mu_N$ being the nuclear dipole moment Bohr magneton, respectively. 
The single-particle expression for the $\textbf{T}$ operator is given elsewhere \cite{cheng}.

In RCC, the atomic wavefunction $|\Psi_v \rangle$ for a single valence ($v$) open-shell system is given by \cite{lindgren,mukherjee}
\begin{eqnarray}
|\Psi_v \rangle = e^T \{1+S_v\} |\Phi_v \rangle, \label{eqn3}
\end{eqnarray}
where $|\Phi_v \rangle= a_v^{\dagger}|\Phi_0\rangle$, with $|\Phi_0\rangle$ being the Dirac-Fock (DF) state for the closed-shell system.
In the singles and doubles approximation, we have
\begin{eqnarray}
T = T_1 + T_2 
  = \sum_{a,p}a_p^+a_a t_a^p + \frac {1}{4}\sum_{ab,pq}a_p^+a_q^+a_ba_a t^{pq}_{ab} \ \ \ \ \ \label{eqn4} \\ 
S_v =  S_{1v} + S_{2v} 
    = \sum_{p \ne v}a_p^+a_v s_v^p + \frac {1}{2}\sum_{b,pq}a_p^+a_q^+a_ba_v s^{pq}_{vb} , \ \ \ \ \label{eqn5}
\end{eqnarray}
\noindent
where $a$ and $a^{\dagger}$ are the creation and annihilation operators. $a,b,c..$ and $p,q,r..$ represent core and virtual orbitals, respectively. $t_a^p$ and $t^{pq}_{ab}$ are the cluster amplitudes corresponding to the single ($T_1$) and double ($T_2$) excitation operators in the closed-shell and $s_v^p$ and $s^{pq}_{vb}$ are the cluster amplitudes for the single ($S_{1v}$) and double ($S_{2v}$) excitation operators involving the valence electron ($v$).
Equations to determine these amplitudes with partial triple excitation effects are given elsewhere (For example, see \cite{bijaya3,bijaya2}).

The hyperfine constants are calculated using the general expression for an operator $O$ as 
\begin{widetext}
\begin{eqnarray}
 \langle O \rangle_v  &=& \frac {\langle\Psi_v | O | \Psi_v \rangle} {\langle\Psi_v|\Psi_v\rangle} \nonumber \\
 &=& \frac {\langle \Phi_v | \{1+S_v^{\dagger}\} e^{T^{\dagger}} O e^T \{1 + S_v\} | \Phi_v\rangle } {1+N_v} \nonumber \\
 &=& \frac {\langle \Phi_v | \{1+S_v^{\dagger}\} \overline{O} \{1 + S_v\} | \Phi_v\rangle } {1+N_v} \nonumber \\
 &=& \frac {1} {1+N_v}[ {\langle \Phi_0 | \overline{O} | \Phi_0\rangle + 2 \{ \overline{O} S_{1v} + \overline{O} S_{2v} \} + S_{1v}^{\dagger}  \overline{O} S_{1v} +S_{2v}^{\dagger} \overline{O} S_{1v}+S_{2v}^{\dagger} \overline{O}S_{2v} | \Phi_v\rangle } ], \label{eqn6}
\end{eqnarray}
\end{widetext}
where we define the effective operators $\overline{O} = e^{T^{\dagger}} O e^T$ and $N_v = \langle \Phi_v | S_v^{\dagger} e^{T^{\dagger}} e^T S_v|\Phi_v\rangle $. The above expression is calculated using the method described by \cite{bijayaba,bijaya2}. Contributions from the normalization factor has been considered by
\begin{eqnarray}
Norm. = \langle \Psi_v | O | \Psi_v \rangle \{ \frac {1}{1+N_v} - 1 \}. \label{eqn7}
\end{eqnarray}

\begin{table}
\caption{Total number of GTOs and active orbitals considered in the DF and RCC wavefunction calculations.}
\begin{ruledtabular}
\begin{tabular}{lccccccccc}
      & s$_{1/2}$ & p$_{1/2}$ & p$_{3/2}$ & d$_{3/2}$ & d$_{5/2}$ & f$_{5/2}$ & f$_{7/2}$ & g$_{7/2}$ & g$_{9/2}$ \\
\hline
Be$^+$ &  &  &  &  &  &  &  &  &  \\
No. of GTOs & 30 & 25 & 25 & 20 & 20 & 15 & 15 & 15 & 15 \\
Active orbitals & 12  & 11 & 11 & 11 & 11 & 7 & 7 & 5 & 5 \\
Mg$^+$ &  &  &  &  &  &  &  &  &  \\
No. of GTOs & 30 & 25 & 25 & 20 & 20 & 15 & 15 & 15 & 15 \\
Active orbitals & 12  & 11 & 11 & 11 & 11 & 7 & 7 & 5 & 5 \\
Ca$^+$ &  &  &  &  &  &  &  &  &  \\
No. of GTOs & 30 & 25 & 25 & 25 & 25 & 20 & 20 & 20 & 20 \\
Active orbitals & 13  & 12 & 12 & 12 & 12 & 7 & 7 & 5 & 5 \\
Sr$^+$ &  &  &  &  &  &  &  &  &  \\
No. of GTOs & 35 & 30 & 30 & 25 & 25 & 20 & 20 & 20 & 20 \\
Active orbitals & 13  & 12 & 12 & 12 & 12 & 7 & 7 & 5 & 5 \\
Ba$^+$ &  &  &  &  &  &  &  &  &  \\
No. of GTOs & 38 & 35 & 35 & 30 & 30 & 25 & 25 & 20 & 20 \\
Active orbitals & 13  & 13 & 13 & 13 & 13 & 9 & 9 & 8 & 8 \\
\end{tabular}
\end{ruledtabular}
\label{tab:d5hyp1}
\end{table}
\begin{table}[t]
\caption{$g_I$ values used in the present work. }
\begin{ruledtabular}
\begin{center}
\begin{tabular}{lr@{.}lr@{.}lr@{.}lr@{.}lr@{.}lr@{.}lr@{.}lr@{.}l}
 & \multicolumn{2}{c}{Be} & \multicolumn{2}{c}{Mg} & \multicolumn{2}{c}{Ca} & \multicolumn{2}{c}{Sr} & \multicolumn{2}{c}{Ba} \\
\hline \\
$g_I$ & $-$0 &78499 & $-$0&34218 & $-$0&37647 & $-$0&243023 & 0&6249 \\
\end{tabular}
\end{center}
\end{ruledtabular}
\label{tab:d5hyp2}
\end{table}

We use the Dirac-Coulomb atomic Hamiltonian for the present calculations. We employ Gaussian type orbitals (GTOs) to construct the DF wavefunction as explained in \cite{rajat}. Finite nuclei with Fermi charge distribution are considered as given by Parpia and Mohanty \cite{parpia}. All core-electron effects are considered in the DF and RCC wavefunction calculations for all the systems. In table \ref{tab:d5hyp1}, we present the total number of GTOs used for the DF and active orbitals used for the RCC calculations in different systems. In table \ref{tab:d5hyp2}, we present the $g_I$ values that are used to calculate $A$ for various systems.
\begin{table}
\caption{Dirac-Fock, all order CP and experimental results of the magnetic dipole hyperfine constant ($A$) of the $^2D_{5/2}$ states of the alkaline earth ions in MHz.}
\begin{tabular}{lr@{.}lr@{.}lr@{.}lr@{.}lr@{.}l}
\hline
\hline
 & \multicolumn{2}{c}{Be$^+$} & \multicolumn{2}{c}{Mg$^+$} & \multicolumn{2}{c}{Ca$^+$} & \multicolumn{2}{c}{Sr$^+$} & \multicolumn{2}{c}{Ba$^+$} \\
 & \multicolumn{2}{c}{($3d_{5/2}$)} & \multicolumn{2}{c}{($3d_{5/2}$)} & \multicolumn{2}{c}{($3d_{5/2}$)} & \multicolumn{2}{c}{($4d_{5/2}$)} & \multicolumn{2}{c}{($5d_{5/2}$)} \\
\hline \\
DF & $-$1&020 & $-$0&539 & $-$14&163 & $-$13&006 & 53&213 \\
$\overline{O}S_{2v}$ & 0&057 & 0&659 & 15&793 & 18&795 & $-$78&273 \\
Expt. &  \multicolumn{5}{r@{.}}{$-$3}&8(6)$^a$ & 2&1743(14)$^b$ & $-$11&9(10)$^c$ \\
 \multicolumn{10}{r@{.}}{$-$7}&4(10)$^d$ \\
 \multicolumn{10}{r@{.}}{$-$12}&028(11)$^e$ \\
\hline
\hline
\end{tabular}
References: $^a$ \cite{norterhauser}; $^b$ \cite{barwood1}; $^c$ \cite{huennekens}; $^d$ \cite{silverans1}; $^e$ \cite{silverans}.
\label{tab:d5hyp3}
\end{table}
\begin{figure}
\includegraphics[width=6.0cm]{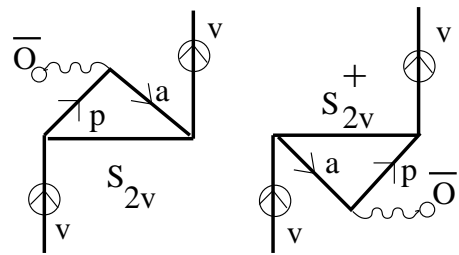}
\caption{Goldstone diagrams representing all order CP effects from $\overline{O}S_{2v}$. Arrows pointing downwards and upwards represent core (a) and virtual (p) electrons, respectively. Arrow inside the circle represents valence electron.}
\label{fig:fig1}
\end{figure}

The terms involving $OS_{2v}$ in the expansion of Eqn. \ref{eqn6} represents the all order CP effect. One could, therefore, consider $\overline{O}S_{2v}$ as an all order dressed CP effect and diagrams associated with them are given in Fig. \ref{fig:fig1}. In order to highlight the spectacular role of the CP effects, we present the DF and $\overline{O}S_{2v}$ contributions along with the experimental results of $A$ for different states in table \ref{tab:d5hyp3}. It is obvious from this table that the DF and experimental results have opposite signs except for Ca$^+$. The contributions from $\overline{O}S_{2v}$ are larger than the DF values for all the ions except Be$^+$ which has only one core-orbital. Their inclusion significantly improves the agreement of $A$ with the experimental values for all the ions.

\begin{figure}
\includegraphics[scale=0.52]{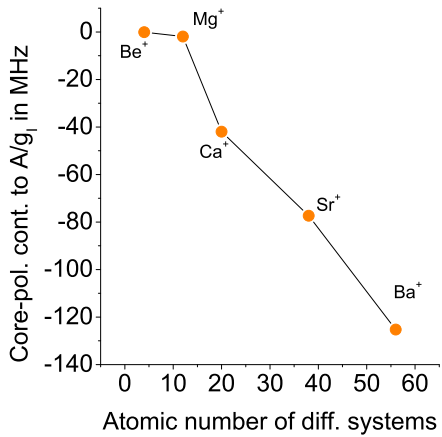} 
\caption{(color online) Core-polarization contributions to $A/g_I$ for different systems.}
\label{fig:fig2}
\end{figure}
A deeper understanding of these results can be obtained from table \ref{tab:d5hyp4}, where we have given the all order CP contribution to $A$ from different core (a) and virtual (p) orbitals. It is clear from this table that the valence or the low-lying continuum orbitals and the outer core orbitals contribute more than the other orbitals. The significant contributions due to the valence orbitals arise from the large size of the hyperfine matrix elements and the small energy differences involving the valence and the outer core orbitals. Even though the high-lying ($s$ and $p$) continuum orbitals  have a large density at the nucleus, their contributions are small. Their energies are rather large compare to the core orbitals and this makes their contributions to $A$ almost insignificant.

From Eqn. \ref{eqn1}, it is clear that $A$ depends on both the nuclear Land\a'e g-factor ($g_I$) and the matrix element of the $\textbf{T}$ operator which depends on the electron wavefunctions. It would, therefore, be appropriate to study the variation in electron correlation in different systems by examining the results of $A/g_I$.

\begin{table}
\caption{CP contributions from single particle orbitals.}
\begin{tabular}{lccccccc}
\hline
\hline
& Core & Virtual & CP & & Core & Virtual & CP \\
 & (a)  & (p) & ($\overline{O}S_{2v}$) & & (a) & (p) & ($\overline{O}S_{2v}$) \\
\hline
Be$^+$ & & & & & & & \\
& $1s_{1/2}$ & $2s_{1/2}$ & \hfill0.0526 & & $1s_{1/2}$ & $6s_{1/2}$ & \hfill0.0028 \\
& $1s_{1/2}$ & $2s_{1/2}$ & \hfill0.0526 & & $1s_{1/2}$ & $6s_{1/2}$ & \hfill0.0028 \\
& $1s_{1/2}$ & $3s_{1/2}$ & \hfill0.0004 & & $1s_{1/2}$ & $8s_{1/2}$ & \hfill$-$0.0008 \\
& $1s_{1/2}$ & $4s_{1/2}$ & \hfill0.0004 & & $1s_{1/2}$ & $7d_{3/2}$ & \hfill0.0004 \\
Mg$^+$ & & & & & & & \\
& $1s_{1/2}$ & $8s_{1/2}$ & \hfill$-$0.0056 & & $2s_{1/2}$ & $9s_{1/2}$ & \hfill0.0192 \\
& $1s_{1/2}$ & $9s_{1/2}$ & \hfill$-$0.0097 & & $2p_{1/2}$ & $3p_{1/2}$ & \hfill$-$0.0242 \\
& $2s_{1/2}$ & $3s_{1/2}$ & \hfill0.2591 & & $2p_{1/2}$ & $7p_{1/2}$ & \hfill$-$0.0183 \\
& $2s_{1/2}$ & $4s_{1/2}$ & \hfill0.0240 & & $2p_{1/2}$ & $8p_{1/2}$ & \hfill$-$0.0147 \\
& $2s_{1/2}$ & $5s_{1/2}$ & \hfill0.0097 & & $2p_{3/2}$ & $3p_{3/2}$ & \hfill0.1680 \\
& $2s_{1/2}$ & $6s_{1/2}$ & \hfill0.0135 & & $2p_{3/2}$ & $4p_{3/2}$ & \hfill0.0136 \\
& $2s_{1/2}$ & $7s_{1/2}$ & \hfill0.0903 & & $2p_{3/2}$ & $6p_{3/2}$ & \hfill0.0114 \\
& $2s_{1/2}$ & $8s_{1/2}$ & \hfill0.0861 & & $2p_{3/2}$ & $7p_{3/2}$ & \hfill0.0498  \\
Ca$^+$ & & & & &  & & \\
& $2s_{1/2}$ & $8s_{1/2}$ & \hfill0.6576 & & $2p_{1/2}$ & $9p_{1/2}$ & \hfill$-$0.3196 \\
& $2s_{1/2}$ & $9s_{1/2}$ & \hfill1.6782 & & $3p_{1/2}$ & $8p_{1/2}$ & \hfill$-$0.2283 \\
& $2s_{1/2}$ & $10s_{1/2}$ & \hfill0.8119 & & $3p_{1/2}$ & $9p_{1/2}$ & \hfill$-$0.5072 \\
& $3s_{1/2}$ & $4s_{1/2}$ & \hfill2.0285 & & $3p_{3/2}$ & $4p_{3/2}$ & \hfill0.7231 \\
& $3s_{1/2}$ & $7s_{1/2}$ & \hfill0.8361 & & $3p_{3/2}$ & $7p_{3/2}$ & \hfill0.4276 \\
& $3s_{1/2}$ & $8s_{1/2}$ & \hfill3.1017 & & $3p_{3/2}$ & $8p_{3/2}$ & \hfill0.9689 \\
& $3s_{1/2}$ & $9s_{1/2}$ & \hfill2.8690 & & $3p_{3/2}$ & $9p_{3/2}$ & \hfill0.8745 \\
Sr$^+$ & & & & & & & \\
& $2s_{1/2}$ & $11s_{1/2}$ & \hfill0.8428 & & $4p_{1/2}$ & $9p_{1/2}$ & \hfill$-$0.9871 \\
& $4s_{1/2}$ & $5s_{1/2}$ & \hfill4.4052 & & $4p_{1/2}$ & $5p_{3/2}$ & \hfill1.2774 \\
& $4s_{1/2}$ & $6s_{1/2}$ & \hfill0.9198 & & $4p_{3/2}$ & $8p_{3/2}$ & \hfill0.9915 \\
& $4s_{1/2}$ & $8s_{1/2}$ & \hfill2.3399 & & $4p_{3/2}$ & $9p_{3/2}$ & \hfill2.1803 \\
& $4s_{1/2}$ & $9s_{1/2}$ & \hfill6.6872 & & $4p_{3/2}$ & $10p_{3/2}$ & \hfill0.7558 \\
& $4s_{1/2}$ & $10s_{1/2}$ & \hfill$-$0.8056 & & $3d_{5/2}$ & $10d_{5/2}$ & \hfill$-$0.4921 \\
Ba$^+$ & & & & &  & & \\
& $2s_{1/2}$ & $12s_{1/2}$ & \hfill$-$1.9393 & & $5p_{1/2}$ & $9p_{1/2}$ & \hfill2.2198 \\
& $4s_{1/2}$ & $10s_{1/2}$ & \hfill2.2430 & & $5p_{1/2}$ & $10p_{1/2}$ & \hfill4.6907 \\
& $4s_{1/2}$ & $11s_{1/2}$ & \hfill$-$1.8759 & & $3p_{3/2}$ & $13p_{1/2}$ & \hfill$-$1.0486 \\
& $5s_{1/2}$ & $6s_{1/2}$ & \hfill$-$20.5902 & & $4p_{3/2}$ & $10p_{1/2}$ & \hfill1.5367 \\
& $5s_{1/2}$ & $7s_{1/2}$ & \hfill$-$4.5075 & & $5p_{3/2}$ & $6p_{3/2}$ & \hfill$-$4.3027 \\
& $5s_{1/2}$ & $8s_{1/2}$ & \hfill$-$1.5996 & & $5p_{3/2}$ & $7p_{1/2}$ & \hfill$-$1.1738 \\
& $5s_{1/2}$ & $9s_{1/2}$ & \hfill$-$16.6304 & & $5p_{3/2}$ & $9p_{3/2}$ & \hfill$-$5.9189 \\
& $5s_{1/2}$ & $10s_{1/2}$ & \hfill$-$28.6553 & & $5p_{3/2}$ & $10p_{3/2}$ & \hfill$-$8.7231 \\
& $2p_{1/2}$ & $12p_{1/2}$ & \hfill1.0509 & & $4d_{5/2}$ & $10d_{5/2}$ & \hfill1.7734  \\
& $5p_{1/2}$ & $7p_{1/2}$ & \hfill1.2755 & & $4d_{5/2}$ & $11d_{5/2}$ & \hfill1.9755  \\
\hline
\hline
\end{tabular}
\label{tab:d5hyp4}
\end{table}
\begin{table}[h]
\caption{The magnetic dipole hyperfine structure constants ($A$) in MHz. }
\begin{ruledtabular}
\begin{center}
\begin{tabular}{rcr@{.}lr@{.}lr@{.}l}
System  & State & \multicolumn{2}{c}{This work} & \multicolumn{2}{c}{Others} & \multicolumn{2}{c}{Experiment} \\ 
\hline \\
$^9$Be$^+$ & 3d $^2D_{5/2}$ & $-$0&\multicolumn{5}{l}{976} \\
$^{25}$Mg$^+$ & 3d $^2D_{5/2}$ & 0&107 & 0&\multicolumn{3}{l}{1196 \cite{safronova}} \\
$^{43}$Ca$^+$ & 3d $^2D_{5/2}$ & $-$3&351 & $-$5&2 \cite{martensson} & $-$3&8(6)\\
\multicolumn{5}{r@{.}}{$-$4}&2 \cite{martensson1} \\
\multicolumn{5}{r@{.}}{$-$3}&552 \cite{yu} \\
\multicolumn{5}{r@{.}}{$-$4}&84 \cite{itano} \\
$^{87}$Sr$^+$ & 4d $^2D_{5/2}$ & 2&156 & 1&1 \cite{martensson2} & 2&1743(14) \\
\multicolumn{5}{r@{.}}{2}&507 \cite{yu} \\
\multicolumn{5}{r@{.}}{-2}&77 \cite{itano} \\
$^{137}$Ba$^+$ & 5d $^2D_{5/2}$ & $-$11&717 & 9&39 \cite{itano} & $-$12&028(11) \\
\end{tabular}
\end{center}
\end{ruledtabular}
\label{tab:d5hyp5}
\end{table}
\begin{figure}
\includegraphics[scale=0.52]{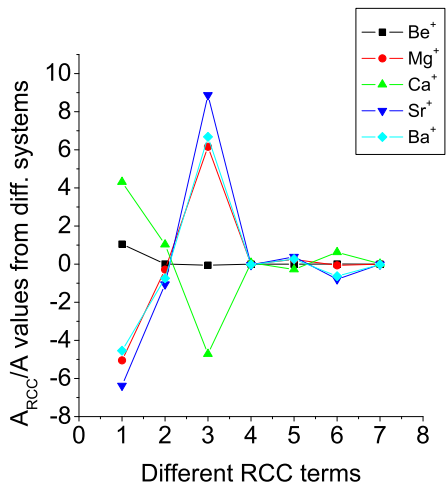}
\caption{(color online) $A_{\text{RCC}}/A$ values for different systems, where the RCC terms correspond to Eqn. \ref{eqn6}: $\overline{O}$ (1), $\overline{O}S_{1v}+cc$ (2), $\overline{O}S_{2v}+cc$ (3), $S_{1v}^{\dagger}\overline{O}S_{1v}$ (4), $S_{2v}^{\dagger}\overline{O}S_{1v}+cc$ (5), $S_{2v}^{\dagger}\overline{O}S_{2v}$ (6) and normalization (7) terms are taken on X- axis.}
\label{fig:fig3}
\end{figure}
The CP effects arising from different systems for the low-lying $^2D_{5/2}$ states are presented in Fig. \ref{fig:fig2} by plotting the ratio of the contributions of $\overline{O}S_{2v}$ with the corresponding $g_I$. These contributions have negative signs and they are very pronounced in heavier systems.

In table \ref{tab:d5hyp5}, we give the results of our RCC calculations for the final values of $A$ as well as those of other calculations and measurements wherever available. In order to appreciate the relative importance of the different physical effects, we plot their fractional contributions in Fig. \ref{fig:fig3} for all the ions. The role played by the different physical effects is evident from this figure. The overall trends exhibited by $A$ for the $^2D_{5/2}$ states of the alkaline earth ions are indeed very peculiar compare to single d- valence neutral atoms \cite{bijaya2} and also single $s$ and $p$ valence atomic systems \cite{bijayaba,bijayapb}. This peculiarity can undoubtedly be attributed to the overwhelming contributions of the CP effects to the hyperfine constants of the states that we have investigated. 

 Our results are considerable improvements over those of previous calculations which were performed by different variants of many-body perturbation theories except the latest which are based on relativistic configuration interaction (RCI) method \cite{itano}. The RCI calculations by Itano use a multi-configuration DF extended optimized level (MCDF-EOL) single particle basis. He considers only a subset of the single and double excitations used in our calculations and the core excitations are limited only to a few outer core electrons. Also, he has excluded the non-linear terms present in our RCC wavefunctions given in Eqn. \ref{eqn3}. For Sr$^+$ and Ba$^+$, the RCI calculations give wrong signs for A (see table \ref{tab:d5hyp5}), thereby highlighting the extraordinarily strong correlation effects in these two systems.

Our present work on the magnetic dipole hyperfine structure constants of the $^2D_{5/2}$ states of the alkaline earth ions is a testament to the remarkable ability of the relativistic coupled-cluster theory to successfully capture very strong correlation effects even when other widely used methods like the finite order relativistic many-body perturbation theory and the relativistic configuration interaction fail. This feature, if suitably exploited, can yield a wealth of very useful information about a wide range of the atomic properties. 

This work was supported by DAAD under the Sandwich Model program (No. A/04/08500).  The authors are grateful to Dr. Wayne Itano for discussions. A part of the computations were carried out on the Teraflop Supercomputer, C-DAC, Bangalore, India.

\end{document}